%% file: main.tex
\documentclass[conference,10pt]{IEEEtran}
\IEEEoverridecommandlockouts

\usepackage{cite}
\usepackage{amsmath,amssymb,amsfonts}
\usepackage{algorithmic}
\usepackage{graphicx}
\usepackage{textcomp}
\usepackage{xcolor}

\usepackage{diagbox}   
\usepackage{pifont}
\usepackage{tabularx}   
\usepackage{array}      

\usepackage{booktabs}
\usepackage{multirow}
\usepackage{graphicx}

\usepackage{amsmath}

\def\BibTeX{{\rm B\kern-.05em{\sc i\kern-.025em b}\kern-.08em
    T\kern-.1667em\lower.7ex\hbox{E}\kern-.125emX}}
\begin{document}

\title{SnakeCharmer: Orchestrating Utility, Timeliness, and Placement in GPU Prefetching
\thanks{\copyright 2026 IEEE. Personal use of this material is permitted. Permission from IEEE must be obtained for all other uses, in any current or future media, including reprinting/republishing this material for advertising or promotional purposes, creating new collective works, for resale or redistribution to servers or lists, or reuse of any copyrighted component of this work in other works.}
}

\input{sec/0-authors}

\maketitle

\input{sec/1-abstract}

\input{sec/2-introduction}

\input{sec/3-background-motivation}

\input{sec/4-design}

\input{sec/5-ex-setup}

\input{sec/6-ex-results}

\input{sec/7-related-work}

\input{sec/8-conclusion}

\bibliographystyle{IEEEtran}
\bibliography{references}

\end{document}

%% file: sec/0-authors.tex
\author{
\IEEEauthorblockN{
Yuhan Tang\IEEEauthorrefmark{2},
Tiejun Li\IEEEauthorrefmark{2}\textsuperscript{*},
Jianmin Zhang\IEEEauthorrefmark{2}\textsuperscript{*},
An Hu\IEEEauthorrefmark{2},
Guangyu Sun\IEEEauthorrefmark{3}, \\
Sheng Ma\IEEEauthorrefmark{2},
Yiqun Lang\IEEEauthorrefmark{2},
Hanqing Li\IEEEauthorrefmark{2},
Siqing Fu\IEEEauthorrefmark{2}
}

\IEEEauthorblockA{
\IEEEauthorrefmark{2}
College of Computer Science and Technology, National University of Defense Technology, China\\
Email: \{tangyuhan, tjli, jmzhang, huan19, masheng, lyq, lihanqing23013, fusiqingnudt\}@nudt.edu.cn
}

\IEEEauthorblockA{
\IEEEauthorrefmark{3}
School of Integrated Circuits, Peking University, China, Email: gsun@pku.edu.cn
}

}

%% file: sec/1-abstract.tex
\begin{abstract}
GPUs rely on massive thread-level parallelism to tolerate long memory latency, yet they still suffer severe stalls when many warps simultaneously wait for data from the memory system. Address-prediction-based GPU prefetchers can mitigate such stalls, but accurate prediction alone does not guarantee performance improvement: a predicted prefetch becomes beneficial only if it is worth issuing, arrives at the right time, and is retained without harmful interference. Because these requirements interact, optimizing any single dimension in isolation can simply move the bottleneck to another.

We capture these requirements with a predicted-prefetch-to-realized-benefit perspective and propose SnakeCharmer, a unified backend prefetch-control framework for address-prediction-based GPU prefetchers. SnakeCharmer separates address prediction from prefetch realization: the frontend predictor generates prefetch candidates, while the backend control path jointly controls utility, timeliness, and placement by determining whether each candidate should be issued, when it should be issued, and where the returned data should be retained. SnakeCharmer realizes this control path with zero-value-aware filtering for utility, hierarchical feedback-driven distance tuning for timeliness, and a register-assisted prefetch buffer for placement. We design and implement SnakeCharmer primarily on top of Snake, and adapt the same backend-control principles to APOGEE to demonstrate that SnakeCharmer's benefits are not limited to a single frontend predictor. On an A100 GPU model, SnakeCharmer improves performance/energy efficiency by 12.8\%/20.8\% over Snake and by 1.7\%/3.2\% over APOGEE, with modest hardware overhead.

\end{abstract}


%% file: sec/2-introduction.tex
\section{Introduction}

Modern GPUs rely on massive thread-level parallelism (TLP) to tolerate long memory access latency~\cite{nematollahi2018neda,chen2014adaptive,darabi2022morpheus,nematollahi2020efficient}. By maintaining many active warps on each streaming multiprocessor (SM), the warp scheduler can switch to another ready warp when the current one stalls on a memory request, thereby overlapping memory latency with useful computation~\cite{narasiman2011improving, damani2022gpu, lee2014caws, lee2015cawa, liu2016barrier}. However, as the gap between GPU compute throughput and memory-system latency continues to widen, TLP alone is increasingly insufficient to fully hide long-latency accesses~\cite{darabi2022morpheus, xu2023ata, arunkumar2017mcm, jing2016cache, ausavarungnirun2015exploiting}. When many resident warps simultaneously wait for data from the L2 cache or off-chip memory, the SM can run out of ready warps and incur substantial idle cycles, leading to underutilized compute resources and degraded performance~\cite{vijaykumar2015case, chaturvedi2024ghost, barnes2023mitigating, iliakis2021repurposing,panda2016prefetching}.

To mitigate such exposed memory latency, prior work~\cite{mostofi2023snake, sethia2013apogee,lee2010many} has explored address-prediction-based GPU hardware prefetching. These prefetchers typically have a logically separable frontend/backend organization: a frontend predictor generates prefetch-address candidates by exploiting regular relationships among memory addresses, while a backend control path determines how these candidates are issued and managed. This scope differs from scheduling-aware prefetchers~\cite{jog2013orchestrated,oh2016apres,koo2018cta} whose candidate generation is tightly coupled with warp/CTA scheduling decisions. However, while accurate address prediction is necessary for prefetching, it is not sufficient to guarantee performance improvement, especially in GPUs where many warps concurrently contend for limited cache capacity, MSHR entries, queues, bandwidth, and on-chip buffering resources. As a result, the benefit of a predicted prefetch depends not only on whether a future address can be identified, but also on whether the candidate is worth issuing, when it should be issued, and where the returned data should be retained. We refer to these requirements as \emph{utility}, \emph{timeliness}, and \emph{placement}. In this work, utility captures whether a candidate is likely to fetch non-zero data rather than low-utility all-zero data; timeliness captures whether the data arrives before, but not too long before, its demand use; and placement captures whether the data can be retained without displacing demand-fetched cache blocks.

Although prior address-prediction-based GPU prefetchers incorporate mechanisms related to these requirements, they typically optimize only one aspect~\cite{mostofi2023snake, sethia2013apogee,lee2010many}. This partial treatment is insufficient because utility, timeliness, and placement interact. Utility alone cannot replace timeliness control, because even a candidate worth issuing provides little benefit if it arrives too early or too late relative to its demand use. Timeliness alone cannot replace placement management, because even a well-timed prefetch may displace demand-fetched data if inserted into a highly contended cache. Placement alone cannot replace utility filtering, because low-utility prefetches may already consume bandwidth, MSHR entries, queues, and on-chip buffering resources before their data is retained. Thus, optimizing only one dimension can simply move the bottleneck to another. This motivates a \textit{predicted-prefetch-to-realized-benefit} perspective, as illustrated in Fig. ~\ref{fig-prefetch-chain}, which views prefetching as a conversion process in which predicted prefetches must satisfy utility, timeliness, and placement requirements before they become realized performance benefits.

\input{fig-prefetch-benefit-chain}

Motivated by this perspective, we propose \textbf{\textit{SnakeCharmer}}, a predictor-agnostic backend prefetch-control framework for address-prediction-based GPU prefetchers. SnakeCharmer separates address prediction from prefetch realization: the frontend predictor generates candidates, while the backend control path determines whether, when, and where each candidate should be handled so that it can turn into a realized benefit. Because SnakeCharmer operates on prefetch candidates rather than on a particular prediction algorithm, it can be integrated with other address-prediction-based prefetchers through only lightweight frontend-specific adaptations, such as candidate metadata and the prefetch-generation interface. We instantiate SnakeCharmer primarily on Snake~\cite{mostofi2023snake} and further demonstrate its portability by adapting it to APOGEE~\cite{sethia2013apogee}.

SnakeCharmer implements this backend with three mechanisms. To improve utility, zero-value-aware filtering suppresses prefetches likely to fetch all-zero data before they consume memory-system resources. To improve timeliness, a hierarchical feedback-driven distance tuner adjusts prefetch distance according to warp execution progress. To improve placement, a register-assisted prefetch buffer opportunistically repurposes idle register capacity to retain prefetched data outside the highly contended L1 cache. On an A100 GPU model, SnakeCharmer improves performance and energy efficiency by 12.8\%/20.8\% over Snake and 1.7\%/3.2\% over APOGEE, showing that its benefits are not limited to a single frontend predictor. This paper makes the following contributions:

\begin{itemize}

\item We introduce the predicted-prefetch-to-realized-benefit perspective, showing that predicted GPU prefetches can fail to deliver performance gains due to interacting utility, timeliness, and placement issues.

\item We propose SnakeCharmer, a backend prefetch-control framework for address-prediction-based GPU prefetchers that jointly controls whether prefetches are issued, when they are issued, and where the returned data is retained.

\item We design and implement SnakeCharmer primarily on top of Snake using zero-value-aware filtering, hierarchical feedback-driven distance tuning, and register-assisted prefetch buffering.

\item We demonstrate SnakeCharmer's broader applicability by adapting its backend-control principles to APOGEE with lightweight frontend-specific modifications.

\end{itemize}

%% file: fig-prefetch-benefit-chain.tex
\begin{figure}[htbp]
    \centering
    \includegraphics[width=1\linewidth]{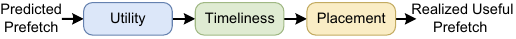}
    \caption{A predicted prefetch becomes a realized benefit only if it satisfies utility, timeliness, and placement requirements.}
    \label{fig-prefetch-chain}
\end{figure}

%% file: sec/3-background-motivation.tex
\section{Background and Motivation}

\subsection{GPU Hardware Prefetching}
Massive warp concurrency is the primary latency-hiding mechanism in GPUs, but it is fundamentally reactive: a demand miss first stalls the issuing warp, and the warp scheduler then attempts to issue another ready warp~\cite{nematollahi2018neda,chen2014adaptive,darabi2022morpheus,nematollahi2020efficient}. Prior work improves this mechanism by better organizing warp scheduling or expanding the effective warp pool, including two-level warp scheduling~\cite{narasiman2011improving}, criticality-aware scheduling~\cite{lee2014caws,lee2015cawa}, virtual threads~\cite{yoon2016virtual}, pre-execution~\cite{kim2016warped}, and lightweight out-of-order execution~\cite{chaturvedi2024ghost}. These techniques reduce exposed memory latency by providing more ready work to the scheduler, but they remain limited when many resident warps simultaneously wait for long-latency memory accesses~\cite{kim2016warped,chatterjee2014managing}. In this regime, thread-level parallelism alone cannot fully hide memory latency, leading to idle execution resources, performance loss, and higher power consumption~\cite{vijaykumar2015case,chaturvedi2024ghost,barnes2023mitigating,iliakis2021repurposing,panda2016prefetching}.

GPU hardware prefetching~\cite{mostofi2023snake,sethia2013apogee,lee2010many,koo2018cta,oh2016apres} complements warp-level latency hiding by acting before a demand miss becomes critical. Instead of waiting for a demand access to stall a warp, a hardware prefetcher issues predictable future memory requests in advance, allowing them to overlap with other warps' execution or otherwise idle memory-system windows. Prefetching can therefore provide additional benefit when future accesses are predictable and warp-level concurrency alone is insufficient to cover long-latency misses. However, prefetching is not always effective: when latency is already hidden by available warps or when access patterns are difficult to predict, it may provide little or even negative benefit.

\subsection{Quantifying the Gap: Utility, Timeliness, and Placement}
\label{sec:snake-prefetcher}

\input{fig-snake-arch}

Prior GPU hardware prefetching has largely focused on improving address predictability.   
Snake~\cite{mostofi2023snake} is a state-of-the-art GPU hardware prefetcher, and its overall architecture is shown in Fig. \ref{fig-snake-arch}. The Head Table, indexed by warp ID, records the most recently executed load PC and its corresponding memory address for each warp, thereby capturing address changes between load instructions. The Tail Table, indexed by load PC, organizes and stores observed stride information, including intra-warp strides, inter-warp strides, inter-PC strides\footnote{Renamed from ``Inter-thread'' in~\cite{mostofi2023snake} to better reflect the chain-based stride detection between different PCs.}, and their training states, to determine whether a stride pattern is stable enough to be used for prefetching. The Throttle Engine controls prefetching aggressiveness based on the available L1 cache space and current memory bandwidth utilization. When prefetching causes cache space to become insufficient or memory bandwidth to become saturated, it pauses or limits the prefetch requests issued by Snake.
Using these structures, Snake supports three prediction patterns: (1) intra-warp prefetching, which predicts addresses across future iterations of the same load within a warp; (2) inter-warp prefetching, which enables leading warps to prefetch data for lagging warps along the same instruction stream; and (3) inter-PC prefetching, which captures correlations across load PCs. These three patterns substantially improve address prediction coverage and accuracy over earlier GPU prefetchers.

However, strong address prediction alone does not guarantee speedup, as in cache-constrained, highly concurrent GPUs, prefetches can increase pollution and contention for limited cache and memory-system resources, which motivates backend control over predicted candidates—whether to issue them, when to issue them, and where to retain the returned data—and we quantify these limitations on Snake to show that the key challenge is converting predicted prefetches into realized performance benefits.

Before presenting these measurements, we define the block granularity used throughout this analysis. Unless otherwise specified, both demand and prefetched blocks are measured at the 32B-sector granularity, the fundamental unit of memory transaction in our GPU hierarchy. A demand block refers to a 32B sector fetched by a demand load miss rather than by a prefetch request. As shown in Table~\ref{tab:gpu_config}, the L1 and L2 caches are sectorized (128B lines divided into four 32B sectors), and all memory requests are split and serviced at this granularity.

\input{fig-background-zero-rate}

\textbf{Utility.} In this paper, we define the utility dimension with respect to whether a predicted prefetch fetches zero-valued data. Specifically, we treat prefetches for zero-valued data as a low-utility case and focus on suppressing such prefetches. This definition is motivated by the fact that, although zero values can be represented compactly in a zero-compressed cache~\cite{islam2009zero,dusser2009zero,liu2025sparsity}, prefetching such data still incurs cache accesses and consumes bandwidth and buffering resources. We first analyze sparsity in both demand and prefetch traffic. As shown in Fig.~\ref{fig-memory-request-and-prefetch}a, zero-value ratios are substantial across multiple granularities, reaching 31.4\% for demand requests and 39.2\% for prefetch requests at 4B granularity. We then directly quantify low-utility prefetching by measuring the fraction of zero-valued blocks in L2 prefetch traffic for sparse applications. Fig.~\ref{fig-memory-request-and-prefetch}b shows that 36.8\% of prefetch traffic is zero-valued at 4B granularity, while the corresponding fraction for demand traffic is 31.9\%. These results show that a non-trivial fraction of predicted prefetches fetch zero-valued data, making zero-valued traffic a concrete and measurable source of low-utility prefetching.

\input{fig-background-effective-prefetch}

\textbf{Timeliness.} Even correctly predicted prefetches may arrive too early or too late to hide latency. We define a prefetch as too late if its demand load occurs before the prefetched block becomes available, and as too early if the block arrives in the L1 cache but is evicted or no longer retained before the demand load. Fig.~\ref{fig-snake-effective-prefetch} shows the timeliness distribution of Snake-generated prefetches that are eventually consumed by demand loads. Among these consumed prefetches, 48.3\% arrive on time, 28.7\% arrive too early, and 23.0\% arrive too late. Prefetches that are never consumed are excluded from this timeliness distribution and are captured separately by the accuracy metric. These results show that even correct predictions often fail to translate into effective latency hiding.

This mistiming is difficult to avoid with a fixed prefetch distance because, on GPUs, the same prefetch distance does not imply the same interval from prefetch issuance to the consuming demand load. This interval depends on intervening computation, memory dependences, warp scheduling, and memory-system contention~\cite{lee2012prefetching,lee2010many,sethia2013apogee,jamilan2022apt}, so a fixed prefetch distance can produce very different timeliness outcomes across applications.

\input{fig-background-prefetch-eviction}

\textbf{Placement.} In a conventional GPU memory hierarchy, prefetched blocks are typically inserted into the L1 cache, where they compete with demand-fetched blocks for limited cache capacity. We quantify this placement-induced interference using prefetch-induced demand evictions. The numerator of this metric is measured at L1 sector granularity and counts replacement events in which the evicted victim sector was originally fetched by a demand load, while the incoming sector is brought by a prefetch request. The denominator is the total number of L1 victim-sector evictions. As shown in Fig.~\ref{fig-prefetch-ratio-eviced-demand}, prefetch-induced demand evictions account for 34.0\% of all L1 victim-sector evictions on average. This indicates that Snake-generated prefetches place substantial pressure on the contended L1 cache, displacing demand data and reducing the realized benefit of prefetching~\cite{jog2013orchestrated}.

\subsection{Motivation for SnakeCharmer}
\label{sec:motivated_for_snakecharmer}

The above analysis shows that better address prediction alone is insufficient for effective GPU prefetching. This motivates SnakeCharmer, which targets three backend-control dimensions that determine realized prefetch effectiveness: utility, timeliness, and placement.

First, predictable addresses do not always imply useful prefetches. In sparse GPU applications, recurring access patterns can repeatedly prefetch zero-valued regions, and deep learning workloads often contain contiguous zero-valued regions produced by activation functions such as ReLU~\cite{shomron2018spatial,rhu2018compressing}. This motivates filtering low-utility candidates before they consume cache, queue, MSHR, and bandwidth resources. Second, fixed prefetch distances are poorly matched to GPU execution because the same logical distance can correspond to different time-to-use intervals under warp scheduling, memory dependences, intervening computation, and memory-system contention. This motivates runtime feedback that adapts prefetch distance to dynamic consumption behavior. Third, placing all prefetched data in the L1 cache increases contention with demand-fetched blocks and can evict useful data before consumption. At the same time, GPU register files often contain idle capacity that cannot improve occupancy due to CTA-level resource granularity, providing an opportunity to buffer prefetched data outside the contended L1 cache.

A GPU application consists of one or more kernels. For each kernel, the CUDA compiler toolchain records in the kernel metadata the number of registers statically allocated per thread, which determines the fixed per-thread register allocation used when the kernel is launched. This value, together with shared-memory usage, thread limits and CTA-slot limits, determines the maximum theoretical CTA residency on each SM. Register capacity improves CTA residency only if all resources required by an additional full CTA are simultaneously available. Thus, leftover registers are useful for occupancy only when there is also sufficient shared memory, thread capacity and CTA-slot capacity. When another resource limits occupancy, or when the remaining registers are fewer than the register demand of the next CTA, the leftover registers cannot admit another CTA and therefore cannot be used to increase residency~\cite{hayes2016orion, yoon2016virtual}. Therefore, underutilized registers are common.

Fig.~\ref{fig-register-ratio} quantifies this effect by reporting the average idle register capacity per kernel for each evaluated application, assuming a 256~KB register file per SM. The figure shows that at least 8~KB of register-file capacity remains idle across all evaluated applications, with an average of 21.0~KB idle. This unused capacity motivates using idle register-file capacity as an auxiliary destination for prefetched data, allowing prefetched blocks to be buffered outside the L1 cache when possible.

\input{fig-background-registers-capacity}

%% file: fig-snake-arch.tex
\begin{figure}[htbp]
    \centering
    \includegraphics[width=0.9\linewidth]{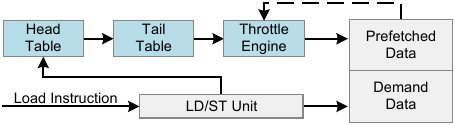}
    \caption{Overview of the Snake prefetcher architecture.}
    \label{fig-snake-arch}
\end{figure}

%% file: fig-background-zero-rate.tex
\begin{figure}[htbp]
    \centering
    \includegraphics[width=1.0\linewidth]{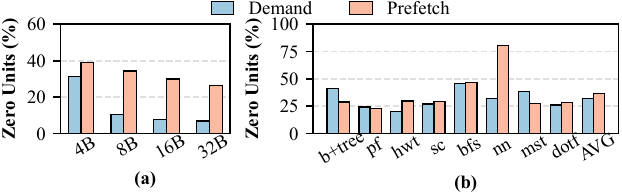}
    \caption{Data sparsity analysis for sparse applications on the A100 architecture using Snake. (a) Zero-valued unit fraction in 32B demand and prefetch blocks partitioned into 4B, 8B, 16B, and 32B units. (b) Zero-valued unit fraction in L2 demand and prefetch traffic at 4B granularity.}
    \label{fig-memory-request-and-prefetch}
\end{figure}

%% file: fig-background-effective-prefetch.tex
\begin{figure}[htbp]
    \centering
    \includegraphics[width=1.0\linewidth]{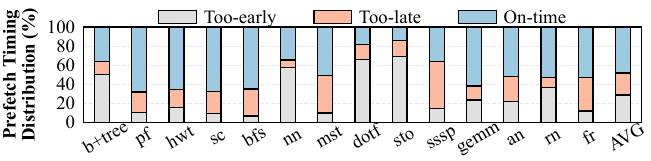}
    \caption{Prefetch timeliness distribution of Snake on the A100 architecture, counting only prefetched blocks that are eventually consumed by demand loads.}
    \label{fig-snake-effective-prefetch}
\end{figure}

%% file: fig-background-prefetch-eviction.tex
\begin{figure}[htbp]
    \centering
    \includegraphics[width=1\linewidth]{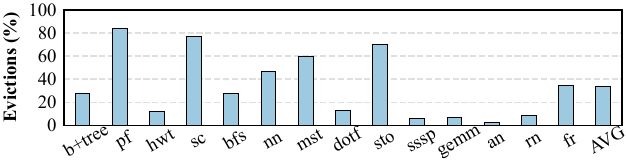}
    \caption{Prefetch-induced demand evictions among all L1 victim-sector evictions on A100 using Snake.}
    \label{fig-prefetch-ratio-eviced-demand}
\end{figure}

%% file: fig-background-registers-capacity.tex
\begin{figure}[htbp]
    \centering
    \includegraphics[width=1\linewidth]{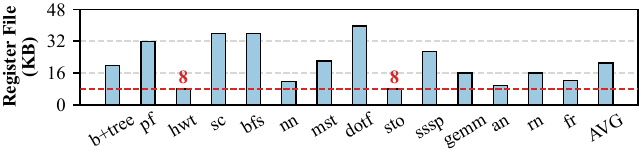}
    \caption{Average idle register file capacity per evaluated application on the A100 architecture.}
    \label{fig-register-ratio}
\end{figure}

%% file: sec/4-design.tex
\section{SnakeCharmer Design}

SnakeCharmer consists of three main components (Fig.~\ref{fig-snakecharmer-arch}), each targeting one backend-control dimension. First, \textit{Snake+} extends the Snake address-prediction frontend with zero-value confidence tracking to filter low-utility candidates. Second, the \textit{SnakeCharmer Controller} (SCC) uses runtime timeliness feedback to tune prefetch distance and gate candidate prefetches. Third, the \textit{Prefetch Remapping Table} (PRT) buffers prefetched data in idle registers, reducing L1-cache interference from prefetched blocks.

SnakeCharmer is designed and evaluated at the maximum warp residency of our GPU configuration, i.e., 64 warps or 2048 threads per SM. Its overhead is fixed per SM rather than per thread: the added structures are fixed-size tables totaling only 2.7 KB per SM (see §\ref{cost_analysis}). Dynamic accesses are bounded by the SM warp-issue and LD/ST pipeline bandwidth, not by the number of resident threads. Prefetch generation and SCC decisions are off the demand critical path; the only demand-side access is the PRT lookup.

\input{fig-snakecharmer-arch}

When a load instruction is executed, the request is forwarded to Snake+ via the Load/Store (LD/ST) unit. Snake+ uses the Head Table and SC-Tail Table to track memory access patterns and generate candidate prefetch requests, while the Throttle Engine preserves Snake's protection against excessive cache and bandwidth pressure. These candidates are then sent to the SCC, which issues them only if they satisfy the confidence and timeliness criteria. If a prefetch is issued, the PRT is probed first; if an available or victim entry exists, the returned data is buffered in idle registers, otherwise it is placed in the L1 cache. On a later demand load, the PRT is checked first, and a hit is served directly from the register file; otherwise, the request proceeds to the L1 cache, where it may hit prefetched data, demand data, or zero-value data managed by the zero-compressed cache. The zero-compressed cache stores compact metadata in a fixed number of L1 cache lines, with each bit indicating whether a corresponding 4B word is all zero.

\subsection{Snake+ Prefetcher}

\input{fig-snake+-prefetch}

Snake+ consists of a Head Table, an SC-Tail Table, and a Throttle Engine. The \textit{Head Table} records, for each warp, the PC of the most recently executed load and its memory access address, indexed by warp ID (WID). On each new load, it updates the corresponding entry and forwards both the previous and updated information to the SC-Tail Table. The \textit{SC-Tail Table}, indexed by PC1 and PC2, stores the detected inter-PC, intra-warp, and inter-warp stride patterns together with their training confidence (T1, T2, T3) and zero-value confidence (Z1, Z2, Z3). The \textit{Throttle Engine} monitors L1 capacity and bandwidth utilization and disables prefetching when resources are insufficient. We empirically configure all candidate prefetch requests to share a single 128-entry prefetch queue.

\textbf{Snake+ without zero-value confidence}. 
Fig.~\ref{fig-snake+-process} illustrates the training and prefetching flow of Snake+ before adding zero-value confidence. For clarity, we assume that one warp is issued at each time step, and $v(t)$ denotes that a value becomes $v$ at time $t$. In this specific example, a prefetch is generated once the corresponding training confidence reaches 3. Snake+ learns and uses three types of stride patterns:
(1) Inter-warp: when consecutive warps execute the same load PC, Snake+ compares their access addresses to learn a stride across warps. For example, warps~1 and~2 access PC~520 at addresses 1000 and 1100, respectively, producing an inter-warp stride of 100; when warp~3 later accesses 1200, T3 is increased and Snake+ prefetches address~1300 for warp~4.  
(2) Inter-PC: when a warp transitions between two load PCs, Snake+ records the address delta as a PC-chain stride. For example, when warp~1 moves from PC~520 at address~1000 to PC~540 at address~1500, Snake+ records a stride of 500 for the chain PC~520$\rightarrow$PC~540. 
(3) Intra-warp and combined prefetching: when warp~1 later returns to PC~520 and accesses address~2500, Snake+ can use the learned inter-PC stride to prefetch $2500+500=3000$, while also using the inter-warp stride to prefetch addresses 2600, 2700, and 2800 for lagging warps~2, 3, and~4. We set the maximum prefetch distances for inter-PC, intra-warp, and inter-warp patterns to 10, 8, and 63, respectively.

\input{fig-snake+-zero}

\textbf{Snake+ with zero-value confidence}. Snake+ uses per-pattern zero-value confidence to filter low-utility prefetches. Each SC-Tail Table entry maintains three 2-bit counters, Z1, Z2, and Z3, for inter-PC, intra-warp, and inter-warp patterns, respectively. Each counter is initialized to 0 and ranges from 0 to 3. A candidate prefetch is suppressed only when the counter associated with its source pattern reaches 3, ensuring that suppression requires repeated zero-value evidence. For an issued prefetch, Snake+ updates the corresponding counter when the candidate block either returns to the L1 cache or already resides in the L1 cache. The counter is incremented if the accessed data is all zero at the tracked access granularity, and decremented otherwise, while remaining bounded within its 2-bit range. Fig.~\ref{fig-snake+-zero} shows an example where the inter-PC candidate is suppressed because Z1=3, while the inter-warp candidate is issued because Z3=0. For a suppressed pattern, Snake+ starts a suppression timer and resets the corresponding zero-value confidence counter to 0 after 128 cycles. We choose 128 cycles because it is sufficient to cover a scheduling pass over the 64 resident warps in our model, thereby avoiding prolonged suppression after the data phase changes. This update-and-timeout policy reduces false suppression and allows fast recovery when the data phase changes.

Zero-value filtering and zero-compressed caching are complementary because they act at different points in the prefetch lifetime. Zero-compressed caching is reactive: it is consulted only after a prefetch request has been issued and has entered the cache hierarchy. As a result, identifying a zero-valued block still requires a cache access and consumes cache-side resources, such as cache ports and cache data bandwidth. Once the cache identifies the requested data as all zero, it can avoid accessing lower cache levels or memory. In contrast, zero-value filtering is proactive: it operates before a candidate prefetch is issued, using zero-value confidence to predict whether the candidate is likely to fetch a zero-valued block. When the confidence is high, SnakeCharmer suppresses the prefetch entirely. Thus, zero-value filtering reduces redundant prefetch requests at their source, avoiding the cache accesses and cache-side resource usage that zero-compressed caching alone cannot eliminate.

\subsection{SnakeCharmer Controller}

As shown in Fig.~\ref{fig-controller}, the SnakeCharmer Controller (SCC) improves prefetch timeliness by adapting prefetch distance from runtime feedback. It consists of a hierarchical Adaptive Window Tuner (AWT) and a Scoring and Gating Unit (SGU). The AWT has an L1 table indexed by SC-Tail rows and an L2 table indexed by prefetch type. L1-AWT tracks on-time, too-early, and too-late outcomes and learns a Local Best Dist after 31 feedback samples; L2-AWT provides a Global Best Dist for cold-start or infrequently executed patterns. Every 15 feedback events, the best distance is updated: it is kept if on-time outcomes exceed 50\%, otherwise shifted by one toward the dominant error.

\input{fig-controller}

As shown in (\ref{equation-sgu}), the SGU uses $S_{\text{total}}$ to gate each candidate, where $T$ is the training confidence and $\Delta D$ is the distance deviation from the selected best distance. Prefetches with higher confidence and smaller deviation are prioritized.

\begin{equation}
\label{equation-sgu}
\small
\begin{aligned}
S_{\text{total}} &= S_{\text{conf}} + S_{\text{dist}}, \qquad
S_{\text{total}} \ge 4 \Rightarrow \text{Issue}, \\
S_{\text{conf}} &=
\begin{cases}
3, & T=7\\
2, & 5\le T\le 6\\
1, & 3\le T\le 4\\
0, & \text{otherwise}
\end{cases}
\qquad
S_{\text{dist}} =
\begin{cases}
3, & \Delta D\le 1\\
2, & 1<\Delta D\le 2\\
1, & 2<\Delta D\le 3\\
0, & \text{otherwise}.
\end{cases}
\end{aligned}
\end{equation}

The Prefetch Eviction Buffer (PEB) is a 64-entry table that records prefetched blocks evicted before first use, enabling SCC to distinguish too-early but eventually demanded prefetches from never-used ones. Training feedback is obtained as follows: a hit in the PRT or L1 cache is on-time, a hit in the PEB is too-early, and a hit in the MSHR is too-late. At prediction time, SCC uses the Local Best Dist when the corresponding L1-AWT entry has warmed up; otherwise, it falls back to the Global Best Dist, and SGU then decides whether to issue the prefetch.

\subsection{Prefetch Remapping Table}

\input{fig-prt}

SnakeCharmer uses a Prefetch Remapping Table (PRT) to buffer prefetched data in idle physical registers rather than always placing it in L1. PRT reserves registers for returning prefetches, allowing later demand loads to read them directly from the register file (RF).

The PRT is a metadata-only lookup table and does not store prefetched data. Prefetched data is stored in reserved idle physical registers in the RF, and all accesses use the existing RF ports. When prefetched data returns to the SM, it is written through the existing LD/ST fill/writeback path, consuming RF write bandwidth; if the target RF bank or write port is busy, the write is queued. Likewise, reads of prefetched data use the RF read ports, consume RF read bandwidth, and are subject to RF bank conflicts. Our simulator explicitly models these RF bank and port conflicts.

Motivated by Fig.~\ref{fig-register-ratio}, which shows that each evaluated application provides at least 8~KB of idle register capacity on average across its kernels, we conservatively configure PRT to use at most 8~KB of idle register capacity, corresponding to 64 PRT entries. SnakeCharmer does not assume that every kernel always provides 8~KB of idle register capacity: if less capacity is available, only a subset of entries is enabled; if none is available, PRT is disabled. A 1-bit enable flag (\texttt{EN}) indicates whether a PRT entry is active. This ensures that PRT uses only registers not used by any active thread.

Each PRT entry contains a tag, four physical base indexes, and four state fields, as shown in Fig.~\ref{fig-prt}. The tag records the memory address of the prefetched data. Each physical base index points to eight consecutive idle physical registers, which together store 32B, matching one memory transaction. The state fields track the status of the prefetched data. Initially, all four slots in each entry are \textsc{INVALID} (00). When a prefetch is issued, the corresponding slot is allocated and transitions to \textsc{WAITING} (01). Following prior work~\cite{oh2019linebacker,lee2015warped}, we assume that physical registers are allocated sequentially and that unallocated registers remain idle throughout the current kernel. When the prefetched data returns, it is written into the reserved registers and the slot transitions to \textsc{READY} (10). After the data is consumed by a demand load, the slot transitions to \textsc{CONSUMED} (11), and the corresponding block is inserted into L1 to preserve normal demand caching behavior. The L1 refill after consumption is not on the critical path of the triggering demand load. The demand load reads the prefetched data from the RF first, and the data is written back to L1 only afterward. Thus, the refill does not delay the triggering demand load, although it still consumes L1 bandwidth in our model. PRT uses round-robin replacement, prioritizing entries with \textsc{CONSUMED} slots; if a \textsc{WAITING} slot is evicted, the returning data is discarded.

The movement of prefetched data through the RF reuses existing datapaths and does not require a new dedicated datapath. Specifically, (1) when prefetched data returns from the memory subsystem to the SM, it is written into reserved registers through the existing LD/ST fill/writeback path; (2) when a demand load consumes prefetched data from the RF, the data is read through the existing RF read path and written back to the destination register through the normal load writeback path; and (3) when consumed prefetched data is sent toward L1, it reuses the existing LD/ST data-movement path from the RF toward the memory hierarchy, rather than requiring a new dedicated RF-to-L1 datapath. Consequently, PRT integration mainly requires lightweight control logic for path selection, state updates, and invalidation.

To maintain coherence, PRT adopts a single-placement policy upon prefetch fill: if a PRT entry is available, the prefetched block is placed in the RF; otherwise, it is placed in L1. Thus, a prefetched block does not exist as two independently valid copies in both PRT and L1. Store requests also probe the PRT tags. If a store address matches a PRT tag, the corresponding slot transitions to the INVALID state, preventing subsequent demand loads from reading stale data from PRT. We model both register-bank conflicts and invalidation overhead in AccelSim, and include both in the evaluation results.

%% file: fig-snakecharmer-arch.tex
\begin{figure}[htbp]
    \centering
    \includegraphics[width=1\linewidth]{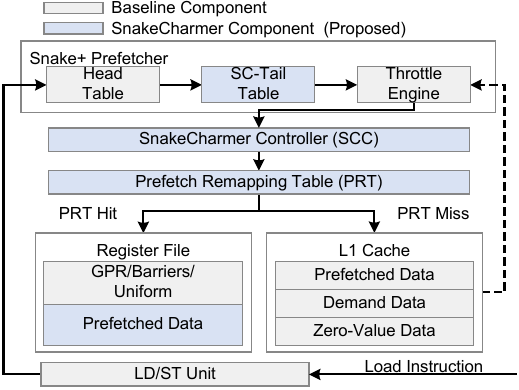}
    \caption{Overview of the SnakeCharmer architecture and the execution flow of load instructions, including the Snake+ Prefetcher, the SnakeCharmer Controller (SCC) and the Prefetch Remapping Table (PRT).}
    \label{fig-snakecharmer-arch}
\end{figure}

%% file: fig-snake+-prefetch.tex
\begin{figure*}[htbp]
    \centering
    \includegraphics[width=1\linewidth]{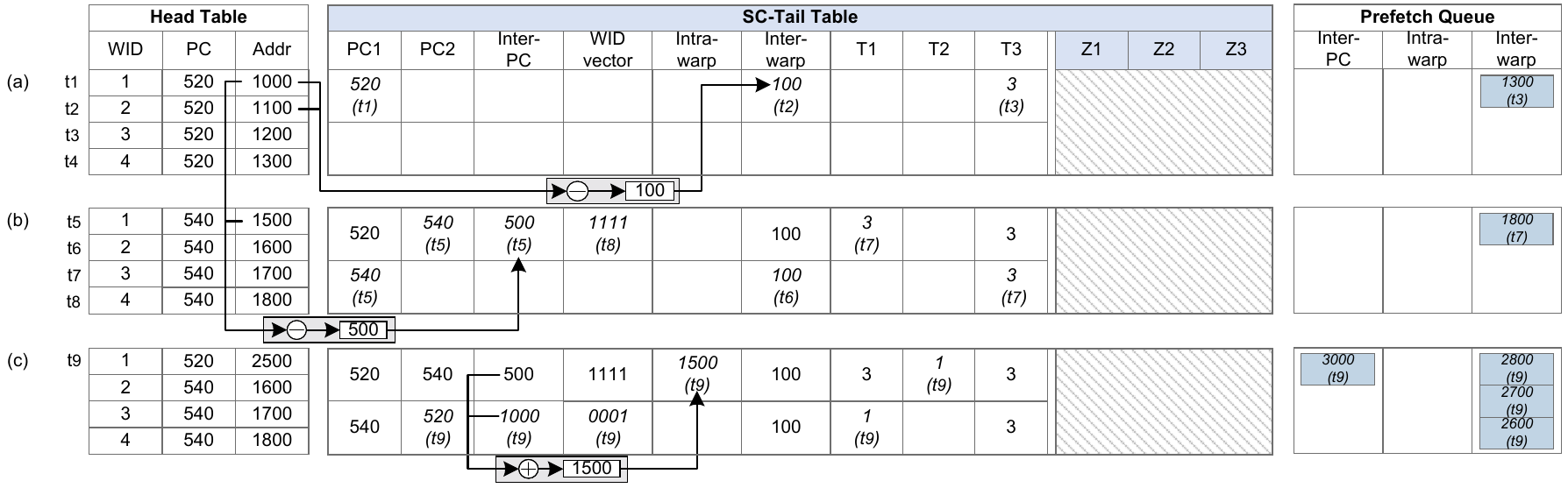}
    \caption{An illustrative example of the training and prefetching workflow of the Snake+ prefetcher, adapted from \cite{mostofi2023snake}. (a) Detection and prefetching of inter-warp strides. (b) Learning of inter-PC stride patterns. (c) Learning of intra-warp strides and combined inter-PC and inter-warp prefetching.}
    \label{fig-snake+-process}
\end{figure*}

%% file: fig-snake+-zero.tex
\begin{figure*}[htbp]
    \centering
    \includegraphics[width=1\linewidth]{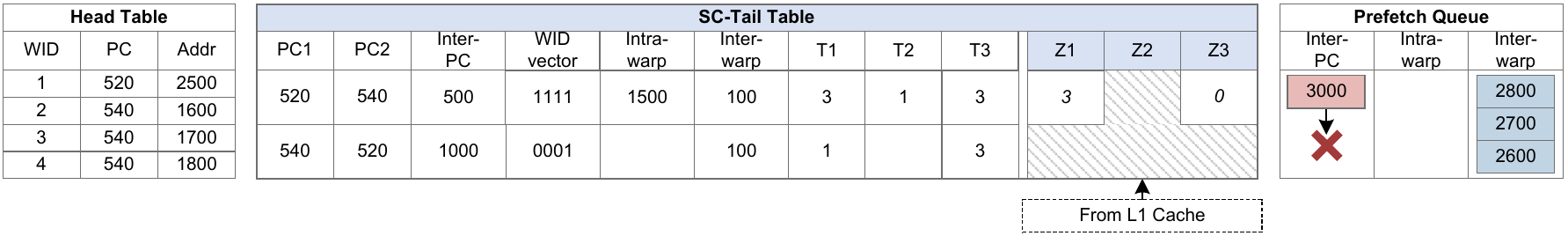}
    \caption{Zero-value confidence-based prefetch decision in the Snake+ prefetcher.}
    \label{fig-snake+-zero}
\end{figure*}

%% file: fig-controller.tex
\begin{figure*}[htbp]
    \centering
    \includegraphics[width=1\linewidth]{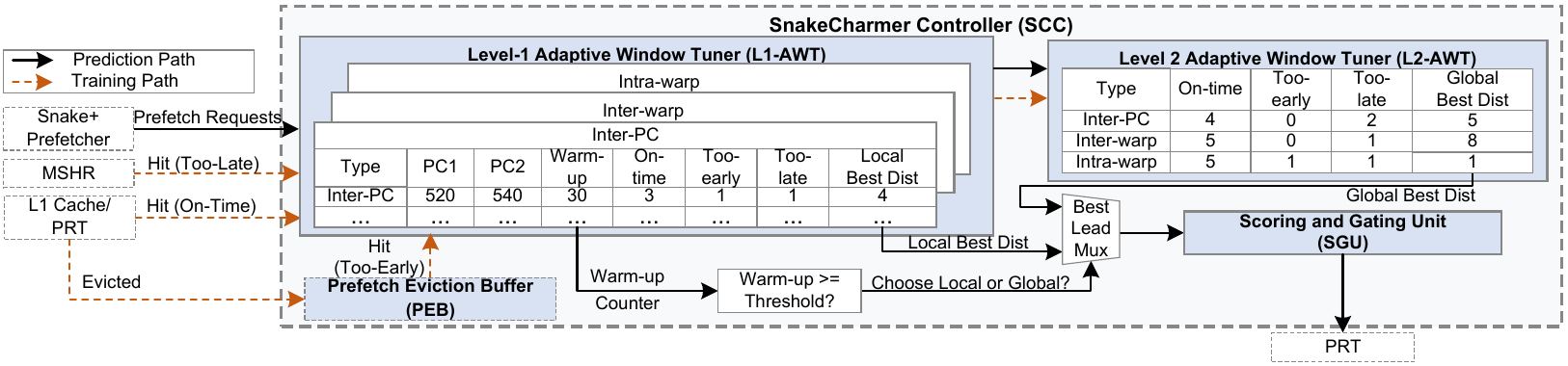}
    \caption{Overall architecture of the SnakeCharmer Controller (SCC). SCC consists of a two-level Adaptive Window Tuner (AWT) hierarchy, a Prefetch Eviction Buffer (PEB), and a Scoring and Gating Unit (SGU) to jointly optimize prefetch timeliness and confidence. Solid and dashed arrows represent prediction and training feedback paths.}
    \label{fig-controller}
\end{figure*}

%% file: fig-prt.tex
\begin{figure}[htbp]
    \centering
    \includegraphics[width=1\linewidth]{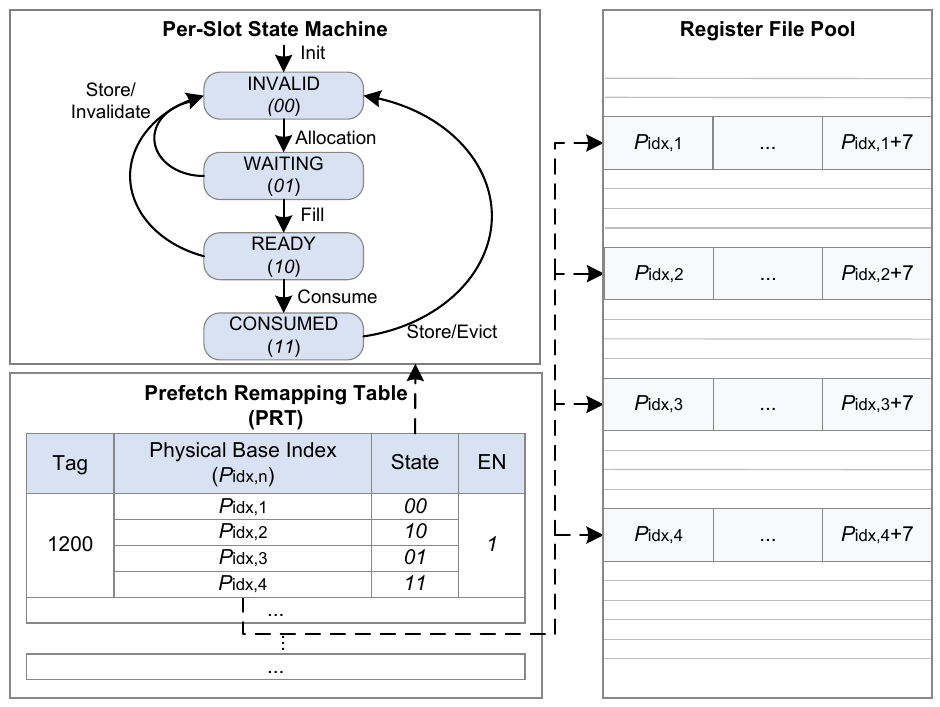}
    \caption{Organization and per-slot state machine of the Prefetch Remapping Table (PRT). }
    \label{fig-prt}
\end{figure}

%% file: sec/5-ex-setup.tex
\section{Experimental Setup}
\label{sec-5}

\textbf{Simulation methodology}. We evaluate SnakeCharmer using the AccelSim cycle-level simulator~\cite{khairy2020accel}. We model NVIDIA A100 and V100 GPUs; unless otherwise specified, all reported results are based on the A100 architecture. Table~\ref{tab:gpu_config} summarizes the parameters of the simulated architectures, and our simulator models the details of the banked register file~\cite{lee2015warped}. To evaluate the power consumption of the overall system, we use AccelWattch~\cite{kandiah2021accelwattch}.

We define the ROI as GPU kernel execution and measure only time spent inside kernels. For each application, the simulator runs all kernels from launch to completion and reports the final aggregate results over these kernels. Because no warmup period is discarded, the reported results include cold-start effects.

\input{tab/tab-3}

\textbf{Applications}. Table~\ref{tab:applications} lists the 14 applications evaluated in this work. We select these applications by first running Snake as the frontend predictor and measuring the ideal prefetch accuracy and ideal prefetch coverage of its prediction stream. We include applications whose ideal prefetch accuracy exceeds 50\% and whose ideal prefetch coverage exceeds 30\%. This criterion reflects the scope of SnakeCharmer: it is designed as a backend for address-prediction-based prefetchers and does not discover new address patterns. If the frontend cannot produce predictions with sufficient accuracy and coverage, the limiting factor is the address predictor rather than SnakeCharmer's utility, timeliness, or placement mechanisms. Focusing on workloads with meaningful address-prediction potential therefore allows us to isolate the contribution of SnakeCharmer's backend control path.
Because zero-value-aware filtering is one component of SnakeCharmer, we further classify applications into high- and low-sparsity groups: applications with more than 20\% zero-valued 4-byte units in demand load responses are categorized as high sparsity, and the rest are categorized as low sparsity.
\input{tab/tab-2}
\textbf{Evaluated Systems}. 
We evaluate ten system configurations. 
(1) \textit{SK}: the baseline Snake prefetcher. 
(2) \textit{SK+ZF}: Snake with only zero-value filtering enabled. 
(3) \textit{SK+SCC}: Snake with only the SCC enabled. 
(4) \textit{SK+PRT}: Snake with only the PRT enabled. 
(5) \textit{SK+ZF+SCC}: Snake with zero-value filtering and the SCC enabled. 
(6) \textit{SK+SCC+PRT}: Snake with the SCC and the PRT enabled. 
(7) \textit{SK+PRT+ZF}: Snake with zero-value filtering and the PRT enabled. 
(8) \textit{APG}: the complete APOGEE prefetcher, including FOA/TIA prefetch generation and APOGEE's original timeliness-control mechanisms. 
(9) \textit{APG+SC}: APOGEE with its FOA/TIA prefetch-generation frontend retained, but with APOGEE's original timeliness-control path replaced by the SnakeCharmer backend. 
(10) \textit{SK+SC}: Snake with the complete SnakeCharmer backend enabled, including zero-value filtering, the SCC, and the PRT. 
We enable the zero-compressed cache for all evaluated configurations.

For \textit{APG+SC}, we use SnakeCharmer as a backend controller for APOGEE-generated prefetch candidates. APOGEE retains its FOA/TIA frontend, which identifies fixed-offset and thread-invariant access patterns and generates candidate prefetch addresses. These candidates, together with prediction metadata such as the load PC, prediction type, warp context, and candidate address, are passed to SnakeCharmer instead of APOGEE's original timeliness-control path. SnakeCharmer then determines whether to issue each candidate, when to issue it, and where to retain the returned data.

%% file: tab/tab-3.tex
\begin{table}[htbp]
\centering
\caption{Baseline GPU Configuration.}
\label{tab:gpu_config}
\begingroup
\renewcommand{\arraystretch}{0.88}
\setlength{\aboverulesep}{0pt}
\setlength{\belowrulesep}{0pt}
\resizebox{\linewidth}{!}{%
\begin{tabular}{lll}
\toprule
& \textbf{A100} & \textbf{V100} \\
\hline
Clock Frequency
& 1410 MHz
& 1132 MHz \\

Number of SMs
& 108
& 80 \\

Max Warps
& 64 (2048 threads)
& 64 (2048 threads) \\

Warp Scheduler
& Round-Robin
& Round-Robin \\

Register File per SM
& \begin{tabular}[c]{@{}l@{}}
    \textcolor[rgb]{0.031,0.031,0.031}{256 KB,~}32 banks, \\
    1R+1W / bank, \\
    256 entries / bank, \\
    256-bit / entry
  \end{tabular}
& \begin{tabular}[c]{@{}l@{}}
    256 KB, 16 banks, \\
    1R+1W / bank, \\
    512 entries / bank, \\
    256-bit / entry
  \end{tabular} \\

\textcolor[rgb]{0.031,0.031,0.031}{Unified}~L1/SMEM per SM
& \begin{tabular}[c]{@{}l@{}}
    \textcolor[rgb]{0.031,0.031,0.031}{192 KB}, 64-way, \\
    128-B line, 32-B sector
  \end{tabular}
& \begin{tabular}[c]{@{}l@{}}
    128 KB, 64-way, \\
    128-B line, 32-B sector
  \end{tabular} \\

L1 MSHR per SM
& 512 entries
& 512 entries \\

L2 Cache
& \begin{tabular}[c]{@{}l@{}}
    \textcolor[rgb]{0.031,0.031,0.031}{40 MB total}, 80 slices, \\
    16-way / slice, 128-B line, \\
    32-B sector
  \end{tabular}
& \begin{tabular}[c]{@{}l@{}}
    6 MB total,  64 slices\\
    24-way / slice, 128-B line, \\
    32-B sector
  \end{tabular} \\

L2 MSHR per slice
& 192 entries
& 192 entries \\
\bottomrule
\end{tabular}%
}
\endgroup
\end{table}

%% file: tab/tab-2.tex
\begin{table}[htbp]
\centering
\caption{Evaluated Applications.}
\label{tab:applications}
\begingroup
\renewcommand{\arraystretch}{0.88}
\setlength{\aboverulesep}{0pt}
\setlength{\belowrulesep}{0pt}
\resizebox{\linewidth}{!}{%
\begin{tabular}{lll}
\toprule
\textbf{Application} & \textbf{Abbr.} & \textbf{Input} \\ \hline
\multicolumn{3}{c}{\textit{High-Sparsity Applications}} \\ \hline
B+ Tree~\cite{che2009rodinia}                          & b+tree & 1M keys, 2 commands \\
Particle Filter\footnotemark~\cite{che2009rodinia}     & pf     & 128 $\times$ 128 $\times$ 10, 1000 particles \\
Heart Wall Tracking~\cite{che2009rodinia}              & hwt    & 48.41 MiB AVI, 20 frames \\
Streamcluster~\cite{che2009rodinia}                    & sc     & 65,536 points $\times$ 256 dims \\
Breadth-First Search~\cite{che2009rodinia}             & bfs    & 1M nodes, 5,999,970 edges \\
Neural Network~\cite{bakhoda2009analyzing}             & nn     & 42,764 records, 5 queries \\
Minimum Spanning Tree~\cite{burtscher2012quantitative} & mst    & 4096 nodes, 59,320 edges \\
Decoder-Only Transformer\footnotemark~\cite{transformerCUDA}
                                                              & dotf   & 32 $\times$ 128 tokens \\ \hline
\multicolumn{3}{c}{\textit{Low-Sparsity Applications}} \\ \hline
StoreGPU~\cite{bakhoda2009analyzing}                   & sto    & 192 KiB buffer \\
Single-Source Shortest Path~\cite{burtscher2012quantitative}
                                                              & sssp   & 1,048,576 nodes, 8,259,994 edges \\
General Matrix--Matrix Multiplication~\cite{grauer2012auto}
                                                              & gemm   & 512 $\times$ 512 $\times$ 512 \\
AlexNet~\cite{karki2019tango}                          & an     & 1000 $\times$ 227 $\times$ 227 $\times$ 3 \\
ResNet~\cite{karki2019tango}                           & rn     & 1 $\times$ 224 $\times$ 224 $\times$ 3 \\
Forman Ricci~\cite{Osama2022EOP,Wang:2017:GGG}
                                                              & fr     & 7135 $\times$ 7135 \\
\bottomrule
\end{tabular}%
}
\endgroup
\end{table}

\footnotetext[2]{We set a fixed random seed to ensure reproducibility of the results.}

\footnotetext[3]{We build a decoder-only Transformer on top of the reference implementation.}

%% file: sec/6-ex-results.tex
\section{Evaluation}
\input{fig-results-acc-cover}

\subsection{Coverage and Accuracy}

Fig.~\ref{fig-coverage-and-accuracy} shows the prefetch coverage and accuracy of the evaluated systems on the A100 architecture. Here, \texttt{Z-AVG} and \texttt{AVG} denote the average coverage and accuracy across high-sparsity applications and all evaluated applications, respectively. We make four observations.

First, \textit{SK+SC} achieves the highest average coverage and accuracy, reaching 32.7\% and 58.8\%, respectively, compared with 11.5\% and 23.1\% for the baseline \textit{SK}. In \textit{SK}, bursty prefetch traffic from massive warp concurrency increases L1 contention and bandwidth pressure, causing many otherwise useful prefetches to be delayed, throttled, or evicted before use. By jointly controlling utility, timeliness, and placement, \textit{SK+SC} substantially improves both metrics.


Second, each SnakeCharmer mechanism is effective in isolation. \textit{SK+ZF} improves average coverage and accuracy to 20.6\% and 40.4\%, respectively (from 11.5\% and 23.1\%), and to 23.7\% and 45.3\% for high-sparsity applications (from 11.7\% and 23.7\%). This shows that zero-value filtering effectively eliminates low-utility prefetch traffic. \textit{SK+SCC} achieves 19.9\% coverage and 40.8\% accuracy by adapting prefetch distance, while \textit{SK+PRT} reaches 16.9\% and 31.6\% by buffering prefetched data in idle registers, thereby reducing L1 contention.

Third, combining any two SnakeCharmer mechanisms consistently yields larger gains than enabling any single mechanism alone. \textit{SK+ZF+SCC}, \textit{SK+SCC+PRT}, and \textit{SK+PRT+ZF} improve average coverage to 27.9\%, 27.1\%, and 26.4\%, respectively, while improving average accuracy to 52.2\%, 44.9\%, and 41.6\%. These results show that optimizing only one backend-control dimension in isolation leaves the others as bottlenecks. Because the three dimensions interact, improving one dimension alone can simply shift the bottleneck elsewhere in the backend control path. SnakeCharmer therefore improves realized prefetch effectiveness by unifying these dimensions within a single backend framework.

Fourth, SnakeCharmer also generalizes beyond the Snake frontend. Relative to \textit{APG}, \textit{APG+SC} improves average coverage from 4.7\% to 7.0\% and average accuracy from 13.7\% to 26.5\%, showing that SnakeCharmer can also improve the realized effectiveness of prefetches generated by a different frontend predictor. The gains, however, remain modest because APOGEE’s FOA/TIA-based frontend exposes a narrower set of prefetch opportunities and cannot generate sufficiently many accurate candidate prefetches. This is consistent with SnakeCharmer’s role as a backend framework: when the frontend cannot provide enough accurate candidates, the dominant limitation remains prefetch-address generation rather than backend control.

\input{fig-results-per-energy}

\subsection{Performance and Energy Efficiency Analysis}  
\label{sec:5-energy-efficiency-analysis}  
  
Performance is the end-to-end manifestation of realized prefetch effectiveness. Fig.~\ref{fig-performance-energy} (top) shows the performance of all evaluated systems on the A100 architecture, normalized to the \textit{SK} baseline. We make four key observations.  
  
First, \textit{SK+SC} achieves the best overall performance, improving instructions per cycle (IPC) by 12.8\% on average across all applications and by 14.8\% across high-sparsity applications. These results show that SnakeCharmer consistently increases the realized value of predicted prefetches, with larger gains when sparsity is higher. 

Second, each proposed mechanism is effective in isolation. \textit{SK+ZF}, \textit{SK+SCC}, and \textit{SK+PRT} improve performance by 6.0\%, 7.0\%, and 3.7\% on average, respectively. In particular, \textit{SK+ZF} improves performance by 7.6\% on high-sparsity applications, confirming that zero-value-aware filtering is especially effective when zero-valued data is frequent. 

Third, SnakeCharmer’s main benefit comes from preventing bottleneck shifting across utility, timeliness, and placement, rather than optimizing any one dimension in isolation. Improving only utility does not eliminate mistimed prefetches, improving only timeliness does not prevent cache contention, and improving only placement does not recover bandwidth wasted on low-utility requests. This interaction is evident in the ablation results: \textit{SK+ZF+SCC}, \textit{SK+SCC+PRT}, and \textit{SK+PRT+ZF} improve performance by 10.6\%, 9.9\%, and 9.1\%, respectively, while the full design further raises the gain to 12.8\%. Thus, SnakeCharmer’s advantage comes from jointly orchestrating all three backend control dimensions under a single framework.

Fourth, SnakeCharmer also generalizes beyond the Snake frontend. Replacing APOGEE’s original backend control path with SnakeCharmer improves performance by 1.7\% on average. The gain is smaller than with Snake because APOGEE’s frontend exposes fewer sufficiently accurate and high-coverage prefetch opportunities to the backend. Nevertheless, the consistent improvement shows that SnakeCharmer is not specific to Snake and can also improve the realized effectiveness of prefetches generated by APOGEE.

To evaluate the impact of SnakeCharmer on overall GPU energy efficiency, we calculate performance per watt, i.e., overall GPU IPC divided by average GPU power consumption, for all systems. Fig.~\ref{fig-performance-energy} (bottom) presents the results, normalized to the energy efficiency of the \textit{SK} baseline.
\textit{SK+SC} improves energy efficiency by 20.8\% on average across all applications and by 24.9\% across high-sparsity applications. In comparison, \textit{SK+ZF}, \textit{SK+SCC}, and \textit{SK+PRT} improve energy efficiency by 9.7\%, 10.9\%, and 6.0\%, respectively, while \textit{SK+ZF+SCC}, \textit{SK+SCC+PRT}, and \textit{SK+PRT+ZF} achieve 16.5\%, 14.9\%, and 14.1\% gains. Replacing APOGEE’s original backend with SnakeCharmer also improves energy efficiency by 3.2\%. These gains come from two reinforcing effects: higher IPC increases useful work per unit energy, while SnakeCharmer’s utility- and timeliness-aware control suppresses unnecessary prefetches and reduces wasted cache, bandwidth, and memory-system activity, thereby lowering power consumption.

\subsection{Mechanism Analysis}

\textbf{Bank Conflicts in Register File}. We evaluate the increase in register-file bank conflicts of \textit{SK+SC} over \textit{SK} on the A100 architecture. As shown in Fig.~\ref{fig-rf-bank}, \textit{SK+SC} increases bank conflicts by 6.1\% on average and by up to 15.2\%. This is because the PRT reuses only idle registers and stores a limited number of prefetched blocks, so it introduces modest additional register-file accesses. These results show that using the PRT to repurpose idle registers only marginally increases bank conflicts.

\input{fig-results-reg-conflict}

\textbf{Timeliness Analysis}. To directly evaluate prefetch timeliness, we measure the distribution of prefetch timeliness for \textit{SK+SC} on the A100 architecture. As shown in Fig.~\ref{fig-snake-effective-prefetch}, among prefetched blocks that are eventually consumed by demand loads, \textit{SK} has only 48.3\% on-time prefetches, with 28.7\% too-early and 23.0\% too-late prefetches. As shown in Fig.~\ref{fig-snakecharmer-effective-prefetch}, \textit{SK+SC} increases the on-time fraction to 88.3\% while reducing both too-early and too-late prefetches. This improvement comes from two factors: reduced contention from low-utility and inaccurate prefetches, and adaptive prefetch-distance adjustment that issues high-confidence prefetches at more appropriate times.

\input{fig-results-effective-prefetch.tex}

\textbf{L2 Cache and Off-Chip Memory Traffic}. Reducing L2 cache and off-chip memory accesses is important because both incur long access latency~\cite{oh2019linebacker} and relatively high dynamic access power~\cite{darabi2022morpheus}. In \textit{SK}, L2/DRAM account for 36.4\% of dynamic power on average. We evaluate (1) L2 cache traffic and (2) off-chip memory traffic of \textit{SK} and \textit{SK+SC} on the A100 architecture, as shown in Fig.~\ref{fig-l2-offchip-traffic}. Compared with \textit{SK}, \textit{SK+SC} reduces L2 cache traffic by 23.6\% and off-chip memory traffic by 15.1\% on average. This reduction is mainly attributed to SnakeCharmer's improved utility and timeliness, which reduce unnecessary prefetches issued by \textit{SK}.

\input{fig-results-traffic-l2-dram}

\textbf{Store-Triggered PRT Invalidations}. We measure the fraction of 32B prefetched blocks placed in the PRT that are invalidated by later stores to the same 32B sector before consumption. Fig.~\ref{fig-prt-invalidation} shows the store-triggered PRT invalidation rate of \textit{SK+SC} on the A100 architecture. The rate remains low across applications, averaging only 2.2\% and reaching at most 5.4\%. Thus, coherence-related invalidations have a negligible impact on SnakeCharmer's performance benefit.

\input{fig-results-prt-invalidation}

\textbf{Zero-Value Filtering Error Analysis}. We quantify the filtering errors of the zero-value confidence mechanism in \textit{SK+SC} on the A100 architecture. A false positive is a non-zero prefetch incorrectly suppressed by the filter, while a false negative is a zero-valued prefetch that is still issued. As shown in Fig.~\ref{fig-zero-value-false}, \textit{SK+SC} has a low false-positive rate of 4.6\%, showing that the conservative 2-bit policy rarely suppresses useful non-zero prefetches. The 20.9\% false-negative rate results from requiring repeated zero-value evidence before disabling a pattern. This conservative bias favors robustness while still filtering many zero-valued requests.

\input{fig-false-positive-negative}

\subsection{Sensitivity Analysis}

\textbf{Zero-Compressed Cache Capacity}. The zero-compressed cache incurs no extra area overhead because it merely reserves a fixed fraction of the existing aggregate L1 and L2 capacity for zero-value metadata, with the L2 fractions calculated across all slices rather than per slice. We choose $\frac{1}{16}L1+\frac{1}{16}L2$, which best balances support for zero-valued blocks and capacity for regular data. As shown in Fig.~\ref{fig-zeocache-config}, smaller configurations limit the benefit of zero-value representation, while larger ones reduce the effective cache capacity available for regular data. Compared with $\frac{1}{32}L1+\frac{1}{32}L2$ and $\frac{1}{8}L1+\frac{1}{8}L2$, our configuration improves performance by 7.5\% and 6.0\%, respectively.

\input{fig-results-zero-cache-conf}

\textbf{Different GPU Architectures}. To demonstrate the generality of SnakeCharmer, Fig.~\ref{fig-v100-performance} evaluates it on the V100 architecture. SnakeCharmer improves performance by 19.0\% on average over Snake on V100, exceeding its gain on A100. The larger gain is driven by V100's more constrained cache hierarchy, including a smaller unified L1/shared memory (128 KB vs. 192 KB) and a much smaller L2 cache (6 MB vs. 40 MB), which makes V100 more vulnerable to cache contention and long-latency memory accesses.

\input{fig-results-v100-perf}

\textbf{SGU Threshold}. The SGU issues a prefetch request only when $S_{total}$ reaches or exceeds a predefined threshold. This threshold controls the tradeoff between aggressive prefetching and timeliness-aware gating: a lower threshold issues more prefetches but may increase cache pollution and bandwidth waste, whereas a higher threshold may suppress useful prefetches. Fig.~\ref{fig-SGU-Threshold} shows the results of the SGU threshold sensitivity analysis. Based on this sensitivity analysis, we use a threshold of 4 for A100 and 5 for V100. Relative to per-application optimal thresholds, these unified settings incur a maximum performance loss of only 2.39\% and 1.58\%, respectively, indicating that the threshold selection is robust.

\input{fig-sgu-threshold}

\textbf{PEB Entries}. The PEB stores metadata for prefetched blocks evicted from the PRT or L1 cache before use and provides feedback to the L1-AWT when they are later demanded. A larger PEB captures more blocks that are evicted too early, but it may also retain stale entries whose feedback no longer reflects current timeliness behavior. Fig. \ref{fig-PEB} shows that performance is maximized with 128 PEB entries on both A100 and V100. However, to balance performance and storage overhead, we conservatively choose 64 PEB entries, which still deliver gains of 9.1\% on A100 and 9.6\% on V100. Compared with the per-application optimal PEB sizes, these unified settings incur a maximum performance loss of only 1.5\% and 0.3\%, respectively, indicating that this design choice is robust.

\input{fig-results-peb-entries}

\textbf{AWT Sampling Window Size}. Fig.~\ref{fig-AWT-samping-window} shows the sensitivity of SnakeCharmer to the AWT sampling window size on A100 and V100. Performance is stable for window sizes between 11 and 15, with less than 1\% variation in average performance and less than 1.5\% variation for any individual application. Based on this result, we use a sampling window size of 15.

\input{fig-results-awt-sample}

\textbf{Register Availability and PRT}. Fig.~\ref{fig-PRT-entries} shows the sensitivity of SnakeCharmer to PRT capacity on A100 and V100. Because PRT entries determine how much idle register capacity can be exposed for prefetch buffering, they serve as a practical proxy for effective register availability. A larger PRT generally improves performance by buffering more prefetched data in idle registers, but the benefit diminishes once the number of PRT entries exceeds the idle register capacity available to an application. Beyond this point, additional PRT entries are disabled. In our evaluation, we conservatively use 64 PRT entries, using at most 8 KB of idle registers. This conservative design point helps avoid overstating SnakeCharmer's benefit on applications with abundant idle register capacity.

\input{fig-results-prt-entries}

\textbf{Cache Capacity and Bandwidth}. We evaluate SnakeCharmer on an A100 architecture under different unified L1/shared memory capacities and DRAM bandwidth settings. SnakeCharmer delivers relatively larger gains at smaller cache capacities and lower bandwidth, where prefetch-induced cache pollution and traffic are more costly. Across all evaluated configurations, however, SnakeCharmer consistently outperforms Snake, achieving at least 12.2\% and 11.2\% performance improvements, respectively.

\input{fig-results-cache-bandwidth}

\subsection{Upper-Bound and Oracle Studies}

We use oracle studies to quantify the remaining optimization opportunities. Note that these oracles are not practical designs; unless otherwise specified, they operate on the same prediction stream generated by the Snake frontend. We evaluate five oracles: (1) \textit{O-ZF}, which perfectly filters zero-block prefetches without suppressing useful ones; (2) \textit{O-TIME}, which schedules each predicted prefetch to arrive just before its next demand use; (3) \textit{O-PLACE}, which provides unlimited space for prefetch placement; (4) \textit{O-SC}, which combines the above three backend oracles; 
and (5) \textit{O-L1}, which assumes that all future global memory accesses hit at the L1 access point. Unlike the other oracles, \textit{O-L1} is not constrained by Snake’s prediction stream and therefore provides an upper bound on the benefit of eliminating global-memory access latency.

\input{fig-results-oracle}

The results are shown in Fig.~\ref{fig-oracle-study}. 
We observe that \textit{SK+ZF} trails \textit{O-ZF} by an absolute performance margin of 3.4\%, 
\textit{SK+SCC} trails \textit{O-TIME} by 4.5\%, 
and \textit{SK+PRT} trails \textit{O-PLACE} by 3.7\%. 
In addition, \textit{SK+SC} exhibits absolute performance gaps of 9.7\% and 19.8\% relative to \textit{O-SC} and \textit{O-L1}, respectively. 
The remaining gaps mainly stem from three factors. 
First, SnakeCharmer uses the Snake prefetcher as its prediction frontend, which cannot generate candidate prefetches that cover all future memory accesses. 
Second, zero-value filtering, distance tuning, and PRT placement are finite hardware mechanisms and therefore cannot achieve oracle utility, timeliness, or placement. 
Third, limited L1 capacity, MSHRs, cache ports, and memory bandwidth still cause resource contention.

\subsection{Cost Analysis}
\label{cost_analysis}
To show that SnakeCharmer's joint optimization of utility, timeliness, and placement is practical, we evaluate its hardware overhead in terms of storage, area, latency and power. We first use CACTI v7.0~\cite{muralimanohar2007optimizing} to estimate latency and power at the 22\,nm technology node. We then use DeepScaleTool~\cite{sarangi2021deepscaletool} to scale these results to the 7\,nm node, corresponding to the fabrication process of the NVIDIA A100 GPU. The scaled results are summarized in Table~\ref{tab:hw_storage_overhead}. Finally, we synthesize the RTL models using the ASAP7 7\,nm predictive standard cell library~\cite{vashishtha2017asap7} to estimate area.
\input{tab/tab-6}

SnakeCharmer adds six storage structures: the Head Table, SC-Tail Table, L1-AWT, L2-AWT, PEB, and PRT. Table~\ref{tab:hw_storage_overhead} reports their overhead. In total, SnakeCharmer adds 2.7~KB per SM, or 0.6\% of A100’s on-chip SRAM capacity, and 0.04~$mm^2$, less than 0.005\% of the 826~$mm^2$ A100 die area~\cite{PowerUp-A100}. At 1410~MHz, all added structures are one-cycle accessible. Prefetch generation and SCC decisions are off the demand critical path, with only a one-cycle PRT lookup on the demand side. SnakeCharmer further reduces GPU power by 6.6\% versus Snake, and this overhead is already reflected in the energy-efficiency results in \S\ref{sec:5-energy-efficiency-analysis}.

%% file: fig-results-acc-cover.tex
\begin{figure*}[htbp]
    \centering
    \includegraphics[width=1\linewidth]{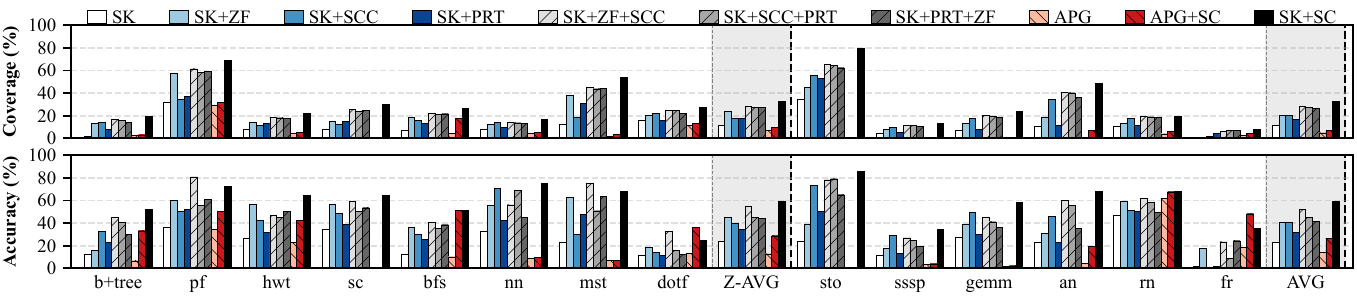}
    \caption{Prefetch coverage (top) and accuracy (bottom) of the evaluated systems on the A100 architecture. Coverage is defined as $H/(H+M)$, representing the fraction of baseline demand misses eliminated by prefetches ($H$ is prefetch hits and $M$ is remaining demand misses). Accuracy is defined as $H/F$, representing the fraction of total prefetch fills ($F$) that are eventually consumed by demand loads.}
    \label{fig-coverage-and-accuracy}
\end{figure*}

%% file: fig-results-per-energy.tex
\begin{figure*}[htbp]
    \centering
    \includegraphics[width=1\linewidth]{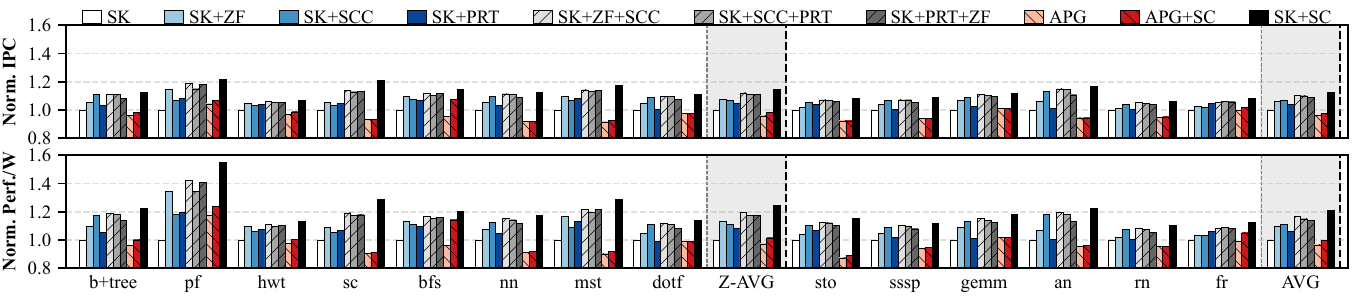}
    \caption{Comparison of IPC (top) and performance per watt (bottom) across all applications on the A100 architecture, normalized to \textit{SK}.}
    \label{fig-performance-energy}
\end{figure*}

%% file: fig-results-reg-conflict.tex
\begin{figure}[htbp]
    \centering
    \includegraphics[width=1\linewidth]{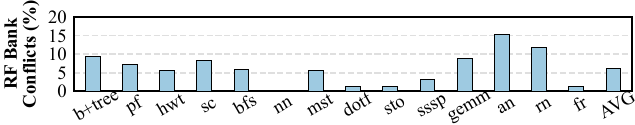}
    \caption{RF Bank-conflict increase of \textit{SK+SC} over \textit{SK} on the A100 architecture.}
    \label{fig-rf-bank}
\end{figure}

%% file: fig-results-effective-prefetch.tex
\begin{figure}[htbp]
    \centering
    \includegraphics[width=1\linewidth]{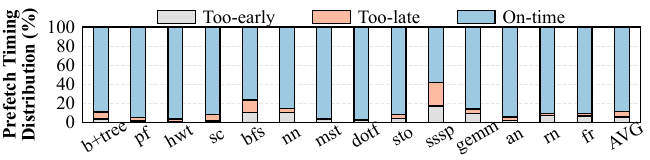}
    \caption{Prefetch timeliness distribution of \textit{SK+SC} on the A100 architecture, counting only prefetched blocks that are eventually consumed by demand loads.}
    \label{fig-snakecharmer-effective-prefetch}
\end{figure}

%% file: fig-results-traffic-l2-dram.tex
\begin{figure}[htbp]
    \centering
    \includegraphics[width=1\linewidth]{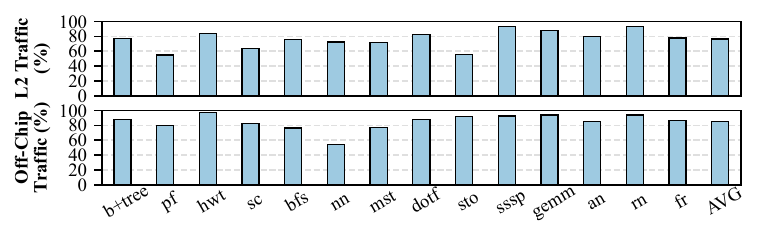}
    \caption{L2 cache and off-chip memory traffic of \textit{SK+SC} on the A100 architecture, normalized to \textit{SK}.}
    \label{fig-l2-offchip-traffic}
\end{figure}

%% file: fig-results-prt-invalidation.tex
\begin{figure}[htbp]
    \centering
    \includegraphics[width=1\linewidth]{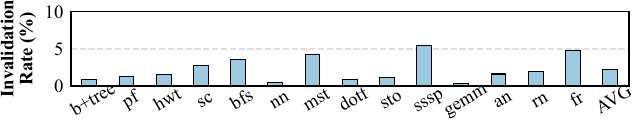}
    \caption{Store-triggered PRT invalidation rate of \textit{SK+SC} on the A100 architecture.}
    \label{fig-prt-invalidation}
\end{figure}

%% file: fig-false-positive-negative.tex
\begin{figure}[htbp]
    \centering
    \includegraphics[width=1\linewidth]{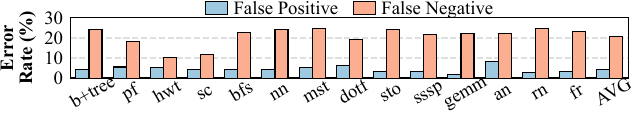}
    \caption{False-positive and false-negative rates of the zero-value confidence mechanism in \textit{SK+SC} on an A100 GPU.}
    \label{fig-zero-value-false}
\end{figure}

%% file: fig-results-zero-cache-conf.tex
\begin{figure}[htbp]
    \centering
    \includegraphics[width=1\linewidth]{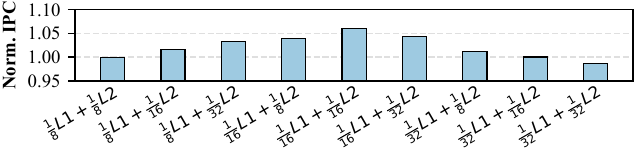}
    \caption{Comparison of different zero-compressed cache configurations. Results are normalized to the 
    $\frac{1}{8}L1 + \frac{1}{8}L2$.}
    \label{fig-zeocache-config}
\end{figure}

%% file: fig-results-v100-perf.tex
\begin{figure}[htbp]
    \centering    
    \includegraphics[width=1\linewidth]{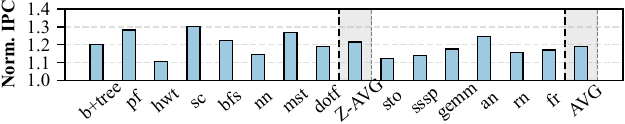}
    \caption{Performance of \textit{SK+SC} on the V100 architecture, normalized to \textit{SK}.}
    \label{fig-v100-performance}
\end{figure}

%% file: fig-sgu-threshold.tex
\begin{figure}[htbp]
    \centering
    \includegraphics[width=1\linewidth]{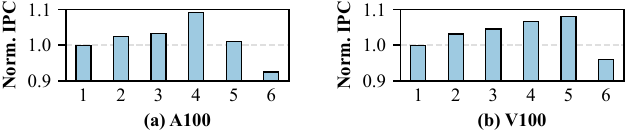}
    \caption{Performance sensitivity to the SGU threshold, normalized to the threshold-1 configuration.}
    
    \label{fig-SGU-Threshold}
\end{figure}

%% file: fig-results-peb-entries.tex
\begin{figure}[htbp]
    \centering
    \includegraphics[width=1\linewidth]{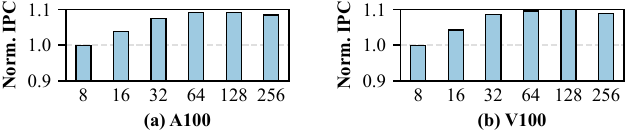}
    \caption{IPC across different PEB sizes, normalized to the 8-entry configuration.}
    
    \label{fig-PEB}
\end{figure}

%% file: fig-results-awt-sample.tex
\begin{figure}[htbp]
    \centering    
    \includegraphics[width=1\linewidth]{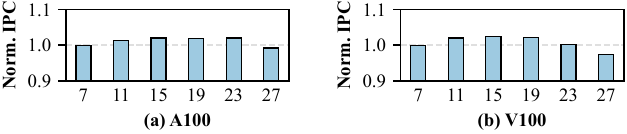}
    \caption{Performance sensitivity to the AWT sampling window size. Results are normalized to the configuration with a window size of 7.}
    \label{fig-AWT-samping-window}
\end{figure}

%% file: fig-results-prt-entries.tex
\begin{figure}[htbp]
    \centering
    \includegraphics[width=1\linewidth]{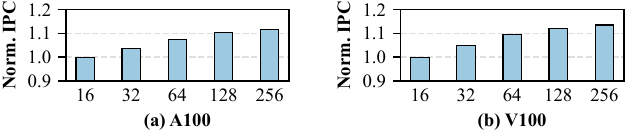}
    \caption{Performance sensitivity to the number of PRT entries. Results are normalized to the 16-entry configuration.}
    
    \label{fig-PRT-entries}
\end{figure}

%% file: fig-results-cache-bandwidth.tex
\begin{figure}[htbp]
    \centering    
    \includegraphics[width=1\linewidth]{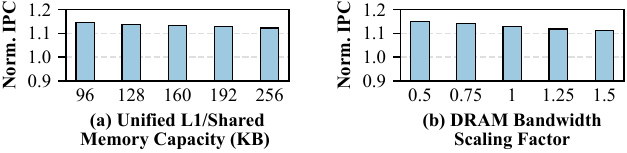}
    \caption{Sensitivity of \textit{SK+SC} to unified L1/shared memory capacity and DRAM bandwidth on the A100 architecture, normalized to \textit{SK}.}
    \label{fig-cache-bandwidth}
\end{figure}

%% file: fig-results-oracle.tex
\begin{figure}[htbp]
    \centering    
    \includegraphics[width=1\linewidth]{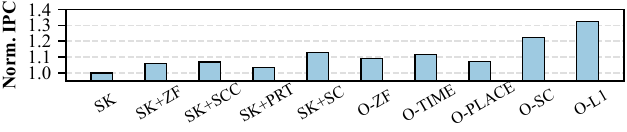}
    \caption{Upper-bound and oracle study results on the A100 architecture, normalized to \textit{SK}.}
    \label{fig-oracle-study}
\end{figure}

%% file: tab/tab-6.tex
\begin{table}[htbp]
\centering
\caption{Storage, access latency, and energy overhead of SnakeCharmer.
Latency and energy values are scaled to the 7\,nm technology node using DeepScaleTool.}
\label{tab:hw_storage_overhead}
\begingroup
\small
\renewcommand{\arraystretch}{0.86}
\setlength{\aboverulesep}{0pt}
\setlength{\belowrulesep}{0pt}
\resizebox{\linewidth}{!}{%
\begin{tabular}{lcccc}
\toprule
\textbf{Component} & \textbf{Configuration} & \textbf{Total} & \textbf{Latency} & \textbf{Energy} \\
& & \textbf{($bytes$)} & \textbf{($ns$)} & \textbf{($pJ$)} \\
\midrule
Head Table    & 10 bytes per entry, 64 entries & 640 & 0.16 & 0.92 \\
SC-Tail Table & 31 bytes per entry, 10 entries & 310 & 0.21 & 3.13 \\
L1-AWT        & 12 bytes per entry, 30 entries & 360 & 0.20 & 0.88 \\
L2-AWT        & 3 bytes per entry, 3 entries   & 9   & 0.09 & 0.08 \\
PEB           & 9 bytes per entry, 64 entries  & 576 & 0.23 & 1.26 \\
PRT           & 13 bytes per entry, 64 entries & 832 & 0.19 & 1.65 \\
\bottomrule
\end{tabular}%
}
\endgroup
\end{table}

%% file: sec/7-related-work.tex
\section{Related Work}

\textbf{Software-Managed Prefetching on NVIDIA GPUs}. Prior work has explored hardware GPU prefetching~\cite{mostofi2023snake,koo2018cta,lakshminarayana2014spare,sethia2013apogee,oh2016apres}, while NVIDIA does not publicly document an L1 hardware prefetcher in its GPUs. NVIDIA instead supports software-managed prefetching, where programmers stage data into shared memory ahead of use~\cite{nvidia_cuda_c}, with recent architectures adding asynchronous global-to-shared copies to better overlap data movement and computation~\cite{nvidia_ampere_whitepaper,nvidia_hopper_whitepaper}. Such techniques require explicit orchestration of data movement, synchronization, and shared-memory capacity, and work best when access patterns or reuse are known ahead of time. SnakeCharmer operates at a different level by transparently controlling hardware-generated prefetch candidates and their placement, so it can target remaining prefetch opportunities not captured by software staging. Consistent with this, \texttt{dotf} and \texttt{sto} already use software-managed prefetching yet still benefit from SnakeCharmer.

\textbf{Scheduling-Aware Hardware Prefetchers}. SnakeCharmer is primarily designed for address-prediction-based GPU prefetchers whose frontend can expose candidate prefetches to a backend control path. Scheduling-aware prefetchers, such as APRES~\cite{oh2016apres}, CAPS~\cite{koo2018cta}, and orchestrated scheduling and prefetching~\cite{jog2013orchestrated}, are less suitable for direct integration because their candidate prefetches are not generated independently of warp/CTA scheduling. For example, APRES forms warp groups based on recently executed load PCs and generates prefetches for other warps in the group after observing the leading warp, CAPS obtains useful candidates only after scheduling leading warps early to compute per-CTA base addresses, and orchestrated scheduling creates candidate prefetches by placing consecutive warps in different scheduling groups so that earlier groups can prefetch for later groups. Applying SnakeCharmer to these designs would therefore require redesigning the scheduler-prefetcher interface, rather than simply replacing a backend prefetch-control path. In this work, we therefore focus on address-prediction-based prefetchers such as Snake and APOGEE, while leaving integration with scheduling-aware prefetchers as future work.

\textbf{Register-Capacity-Based Storage and Prefetching}.    
Prior work has explored using register or local storage capacity for cache-like storage and prefetching~\cite{lakshminarayana2014spare,gebhart2012unifying,oh2019linebacker}. Gebhart et al.~\cite{gebhart2012unifying} and Linebacker~\cite{oh2019linebacker} repurpose local/register-file capacity to increase effective cache-like storage, which can reduce cache pressure and preserve useful lines. These designs, however, primarily target capacity management for demand or victim data; they do not consider the placement of prefetched data or its interaction with prefetch utility and timeliness.

Spare-register-aware prefetching~\cite{lakshminarayana2014spare} uses unallocated registers as destinations for prefetched values in graph workloads and later replaces the corresponding loads with register moves. This mechanism must ensure that the prefetched value matches the value that the later load would have returned before the load can be safely converted into a move. Consequently, it is tied to identifiable loop-based dependent load-pair patterns and does not provide a hardware-managed buffer for address-prediction-based prefetches.
  
These mechanisms show that register/local capacity can provide an alternative placement target, but placement alone is not sufficient: low-utility prefetches can still consume bandwidth and storage, while poorly timed prefetches can still arrive too early or too late. SnakeCharmer therefore treats placement as one component of prefetch realization and coordinates it with utility filtering and timeliness control in a unified backend framework for address-prediction-based GPU prefetchers.

\textbf{CPU Prefetch-Timing Mechanisms}. CPU prefetchers use runtime feedback to adapt prefetch distance~\cite{srinath2007feedback,hur2009feedback}. These mechanisms typically operate on one or a few hardware threads per core, so logical prefetch distance often has a relatively stable relationship with consumption time. This assumption breaks down on GPUs, where many warps are interleaved and dynamically scheduled on each SM; a warp may issue a prefetch, be descheduled, and consume the data much later while other warps make progress. As a result, the same logical distance can produce different real-time lead times across warps and execution phases. SnakeCharmer adapts feedback-driven timing to this GPU setting by associating feedback with warp and load-PC context, allowing the L1/L2-AWT hierarchy to learn distances for individual access patterns rather than relying on a single global setting.

\textbf{GPU Prefetch-Timing Mechanisms.}
MT-prefetching~\cite{lee2010many} throttles prefetching using feedback such as early eviction, while APOGEE \cite{sethia2013apogee} adjusts prefetch distance for FOA/TIA patterns. Both mainly target timeliness. SnakeCharmer treats timeliness as one stage of prefetch realization: its two-level tuner learns pattern-local and global timing behavior and coordinates with zero-value filtering and PRT-based placement. With APOGEE's FOA/TIA frontend unchanged, replacing the full backend improves performance by 1.7\%, showing the effectiveness of the unified backend.

\textbf{Zero-Value Caching.} Zero-value caches compactly represent all-zero blocks to reduce cache pressure and memory traffic~\cite{dusser2009zero,islam2009zero}, and LazyGPU~\cite{liu2025sparsity} combines zero-value caching with lazy execution to eliminate some zero-related memory requests. SnakeCharmer uses zero-compressed caching as common support across all evaluated configurations, following the same zero-value caching idea as LazyGPU. SnakeCharmer's new mechanism is proactive zero-value-aware filtering, which suppresses likely-zero prefetches before they consume queues, MSHRs, bandwidth, or cache resources.

%% file: sec/8-conclusion.tex
\section{Conclusion}

This paper presents SnakeCharmer, a predictor-agnostic backend for address-prediction-based GPU prefetchers. It jointly controls utility, timeliness, and placement through zero-value-aware filtering, feedback-driven distance tuning, and register-assisted buffering, improving performance/energy efficiency by 12.8\%/20.8\% over Snake and 1.7\%/3.2\% over APOGEE on an A100 GPU model.